\documentclass[apj]{emulateapj}
\usepackage{epsfig}

\newcommand{\etal}{et al.}

\newcommand\chandra{{\it Chandra}}
\newcommand\Chandra{{\it Chandra}}
\newcommand\xmm{{\it XMM-Newton}}

\def\psr{\rm{PSR J1852$+$0040}}

\def\one{\rm{1E~1207.4$-$5209}}
\def\pks{\rm{PKS 1209$-$51/52}}

\def\simlt{\mathrel{\hbox{\rlap{\hbox{\lower4pt\hbox{$\sim$}}}\hbox{$<$}}}}
\def\simgt{\mathrel{\hbox{\rlap{\hbox{\lower4pt\hbox{$\sim$}}}\hbox{$>$}}}}

\slugcomment{Submitted to the Astrophysical Journal Letters, April 17 2007}

\shorttitle{Precise Timing of 1E~1207.4$-$5209}
\shortauthors{Gotthelf \& Halpern}

\begin{document}

\title{Precise Timing of the X-ray Pulsar 1E~1207.4$-$5209: \\
A Steady Neutron Star Weakly Magnetized at Birth}

\author{ E. V. Gotthelf \& J. P. Halpern}

\affil{Columbia Astrophysics Laboratory, Columbia University,
New York, NY 10027}

\begin{abstract}
We analyze all X-ray timing data on \one\ in supernova remnant \pks\
gathered in 2000--2005, and find a highly stable rotation with $P =
424.130751(4)$~ms and $\dot P = (9.6 \pm 9.4) \times
10^{-17}$~s~s$^{-1}$. This refutes previous claims of large timing
irregularities in these data.  In the dipole spin-down formalism, the
$2\sigma$ upper limit on $\dot P$ implies an energy loss rate $\dot E
< 1.5 \times 10^{32}$~ergs~s$^{-1}$, surface magnetic field strength
$B_p < 3.5 \times 10^{11}$~G, and characteristic age $\tau_c \equiv
P/2\dot P> 24$~Myr.  This $\tau_c$ exceeds the remnant age by $3$
orders of magnitude, requiring that the pulsar was born spinning at
its present period.  The X-ray luminosity of \one, $L_{\rm bol}
\approx 2 \times 10^{33}\,(d/2\ {\rm kpc})^2$ ergs~s$^{-1}$, exceeds
its $\dot E$, implying that $L_{\rm bol}$ derives from residual
cooling, and perhaps partly from accretion of supernova debris.  The
upper limit on $B_p$ is small enough to favor the electron cyclotron
model for at least one of the prominent absorption lines in its soft
X-ray spectrum.  This is the second demonstrable case of a pulsar born
spinning slowly and with a weak $B$-field, after \psr\ in Kesteven~79.

\end{abstract}


\section {Introduction}

The neutron star \one\ in the center of supernova remnant \pks\ is the
first discovered \citep{hel84} and most intensively studied of the
so-called Central Compact Objects (CCOs). These seemingly isolated NSs
are defined by their steady flux, predominantly thermal X-ray
emission, lack of optical or radio counterparts, and absence of a
surrounding pulsar wind nebula \citep[see][for a review]{pav04}.
\one\ acquired special importance when it became the first CCO in
which pulsations were detected \citep{zav00,pav02}.  It was
distinguished again as the first isolated NS to display strong
absorption lines in its X-ray spectrum \citep{san02,mer02,big03}.

More recently, accumulated X-ray observations of \one\ were presented
as showing large-amplitude changes of both sign in its spin period
\citep{zav04} that were unlike any other pulsar and difficult to
explain.  Consequently, the surface dipole magnetic field, which is a
key parameter in all proposed mechanisms for the X-ray absorption
lines, could not be estimated independently from the spin-down rate,
which was indeterminate.  In this Letter, we present a definitive
study of the spin history of \one\ that corrects previous
errors in the data and their analysis. We provide reliable spin
parameters and discuss their implications for the interpretation of
the X-ray spectrum of \one, and for the origin of the class of CCOs more generally.

\begin{deluxetable*}{cllcccclc}[t]
\tablewidth{0pt}
\tablecaption{Log of X-ray Timing Observations and Summary of Results}
\tablehead{
\colhead{Set}& \colhead{Mission} & \colhead{Instr/Mode}  & \colhead{ObsID/Seq\#} & \colhead{Date} & \colhead{Span} & \colhead{Start Epoch} & 
\colhead{Period\tablenotemark{a}}  & \colhead{$Z^2_1$} \\
\colhead{}        &\colhead{}        & \colhead{}           & \colhead{}            & \colhead{(UT)} & \colhead{(ks)}      & \colhead{(MJD)} & 
\colhead{(ms)}                     & \colhead{}        
}
\startdata
 1& \chandra\ & ACIS-S/CC     & 0751/500249           & 2000 Jan 06 & \phantom{12}32.5           & 51549.625  & 424.13066(48)  &  \phantom{1}51.6  \\
\hline 
 & {\it XMM} & EPIC-pn/SW    & \phantom{/}0113050501 & 2001 Dec 23 & \phantom{12}26.8            & 52266.799  & 424.13075(36)  & 113.1             \\
2& \chandra\ & ACIS-S/CC     & 2799/500249           & 2002 Jan 05 & \phantom{12}30.4            & 52279.952  & 424.13062(38)  &  \phantom{1}62.2  \\
\multispan{4} &  Set 2 combined: &                                                        1167.3 & 52266.799  & 424.130748(16) &  169.6            \\
\hline 
 & {\it XMM} & EPIC-pn/SW    & \phantom{/}0155960301 & 2002 Aug 04 & \phantom{1}128.0            & 52490.306  & 424.130771(41) & 352.6             \\
3& {\it XMM} & EPIC-pn/SW    & \phantom{/}0155960501 & 2002 Aug 06 & \phantom{1}128.4            & 52492.309  & 424.130752(40) & 363.3             \\
\multispan{4} &   Set 3 combined: &                                             \phantom{1}302.0 & 52490.306  & 424.130745(11) & 712.7             \\
\hline 
 & \chandra\ & ACIS-S/CC\tablenotemark{b} & 3915/500394          & 2003 Jun 10 &\phantom{1}155.7 & 52800.443  & 424.13064(13)  &  \phantom{1}26.9  \\
4&\chandra\ & ACIS-S/CC\tablenotemark{b} & 4398/500394           & 2003 Jun 18 &\phantom{1}115.1 & 52808.369  & 424.13084(16)  &  \phantom{1}29.3  \\
\multispan{4} & Set 4 combined: &                                               \phantom{1}799.9 & 52800.443  & 424.130732(12) &  \phantom{1}53.3 \\
\hline 
 & {\it XMM} & EPIC-pn/SW    & \phantom{/}0304531501 & 2005 Jun 22 &            \phantom{12}14.9 & 53543.515  & 424.1299(10)   & \phantom{1}41.7   \\
 & {\it XMM} & EPIC-pn/SW    & \phantom{/}0304531601 & 2005 Jul\phantom{l} 05 & \phantom{12}18.0 & 53556.038  & 424.13078(92)  & \phantom{1}47.5   \\
 & {\it XMM} & EPIC-pn/SW    & \phantom{/}0304531701 & 2005 Jul\phantom{l} 10 & \phantom{12}20.4 & 53561.280  & 424.13127(59)  & \phantom{1}58.7 \\
5 & {\it XMM} & EPIC-pn/SW   & \phantom{/}0304531801 & 2005 Jul\phantom{l} 11 & \phantom{12}63.0 & 53562.090  & 424.13088(15)  &           135.7   \\
 & {\it XMM} & EPIC-pn/SW    & \phantom{/}0304531901 & 2005 Jul\phantom{l} 12 & \phantom{12}13.8 & 53563.283  & 424.13149(75)  & \phantom{1}42.2   \\
 & {\it XMM} & EPIC-pn/SW    & \phantom{/}0304532001 & 2005 Jul\phantom{l} 17 & \phantom{12}16.5 & 53568.016  & 424.13143(66)  & \phantom{1}91.2   \\ 
 & {\it XMM} & EPIC-pn/SW    & \phantom{/}0304532101 & 2005 Jul\phantom{l} 31 & \phantom{12}17.6 & 53582.587  & 424.12910(96)  & \phantom{1}32.3   \\
\multispan{4} & Set 5 combined: &                                                         3393.6 & 53543.515  & 424.1307512(40)& \quad\phantom{2}411.9
\enddata
\tablenotetext{a}{\footnotesize Period derived from a $Z^2_1$ test.
Uncertainty in last digits is in parenthesis, which is
$1\sigma$ computed by the Monte Carlo method described in
\citet{got99}.}
\tablenotetext{b}{\footnotesize These \Chandra\ observations
used the low-energy transmission grating (LETG).}
\label{logtable}
\end{deluxetable*}

\section{Archival X-ray Observations (2000--2005)}

We reanalyzed all timing data on \one\ from the archives of the {\it
Newton X-Ray Multi-Mirror Mission} (\xmm) and \chandra\ observatories.
They span 2000 January to 2005 July.  A log of these observations
is given in Table~\ref{logtable}.

All 11 \xmm\ observations of \one\ used the pn detector of the
European Photon Imaging Camera (EPIC-pn) in ``small window'' (SW)
mode to achieve 5.7~ms time resolution.  Several EPIC-pn data
sets had photon timing errors uncorrected in their original processing
\citep{kir04}.  We reprocessed all EPIC data using the emchain and
epchain scripts under Science Analysis System (SAS) version
xmmsas\_20060628\_1801-7.0.0, which produces correct photon time
assignments.  The observations were affected by background to varying
degree.  To maximize the signal-to-noise ratio in each, we adjusted
the source extraction aperture individually.  For this soft source, an
energy cut of $0.5-2.5$~keV was found to maximize pulsed power.

We also examined data simultaneously available from the EPIC~MOS
camera, operated in ``full frame'' mode.  Although not useful for
timing purposes (2.7~s readout), the location of the source at the
center of the on-axis MOS CCD allows a better background measurement
to test for flux variability, an important indicator of accretion,
than the EPIC-pn SW mode. The seven observations of 2005 exhibit
root-mean-square source flux variability of less than 1\% over the 40
day span.  Comparing these count rates to the earlier \xmm\
observation of 2001 December, implies a marginally significant flux
decrease of $5\% \pm 3\%$ during the 5 year interval.

Four \chandra\ observations suitable for timing measurements of \one\
are available.  They used the Advanced Camera for Imaging and
Spectroscopy (ACIS) in continuous-clocking (CC) mode to provide time
resolution of 2.85~ms.  Two of the four observations were taken with
the LETG transmission grating in place, the zeroth order image being
used for timing. This study uses data processed by the latest pipeline
software (revision v7.6.8.1), with the exception of 2000 January 6,
which is processed with revision v6.5.1.  Reduction and analysis used
the standard software package CIAO (v3.4) and CALDB (v3.3).  The
photon arrival times in CC mode are adjusted in the standard
processing to account for the known position of the pulsar, spacecraft
dither, and SIM offset.  These needed corrections were a potential
cause of timing errors in earlier work, but are now accurately
performed.

\section{Timing Analysis}

\begin{figure}[h]
\centerline{
\hfill
\psfig{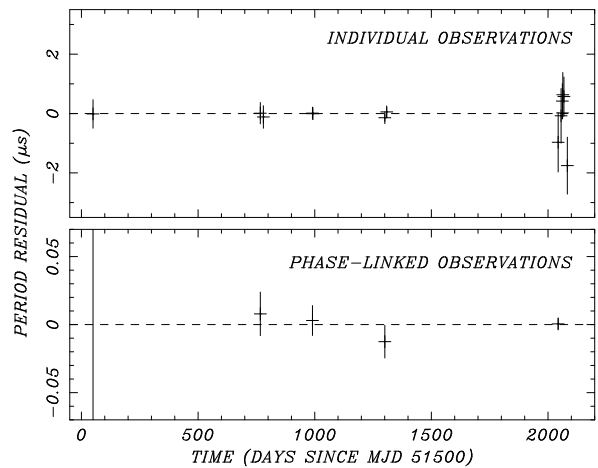}
\hfill
}
\caption{ Period residuals after fitting a linear solution to
individual observations ({\it top panel}) and grouped data sets ({\it
bottom panel}) from Table~\ref{logtable}.  The error bar for the
first individual observation is used, and continues off-scale, in the
bottom panel.}
\label{periodplot}
\end{figure}

For each observation in Table~\ref{logtable} we transformed the photon
arrival times from \one\ to Barycentric Dynamical Time (TDB) using the
coordinates given in Table~\ref{ephemeris}, and identified the pulsed
signal using a standard FFT. To obtain a most precise value of the
period in each observation we then generated a periodogram using a
$Z^2_1$ test \citep{buc83} around the FFT value.  The $1\sigma$
uncertainty in $P$ was determined by the Monte-Carlo method described
in \citet{got99}.  In contrast to the large period changes claimed by
\cite{zav04}, a linear ephemeris is an excellent fit to our derived
periods (see Fig.~\ref{periodplot}), with $\chi^2_{\nu} = 0.65$ for 12
degrees-of-freedom, and no significant detection of a period
derivative.  The average period throughout the data span is $P =
424.130801(57)$~ms, with a formal $2\sigma$ upper limit of $\dot P <
3.9 \times 10^{-15}$~s~s$^{-1}$.

In order to increase the precision, we refitted closely spaced
observations coherently wherever possible.  The results for these
combined data are listed by set in Table~\ref{logtable}, and the
methods are described here. The 2005 June-July set of seven
observations spanned 40 days specifically to obtain a phase-coherent
timing solution.  \citet{woo06} were not able to eliminate large phase
residuals between these observations or find a unique solution. We
determined that the original data processing had timing errors.

\begin{figure}[h]
\centerline{
\hfill
\psfig{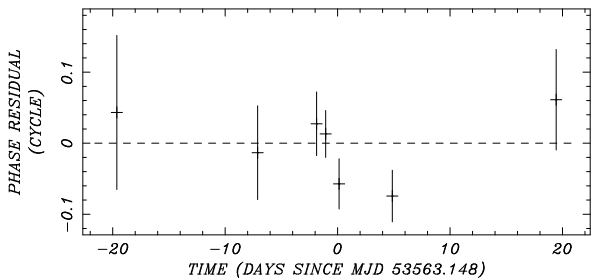}
\hfill
}
\caption{
Pulse phase residuals for the 2005 \xmm\  observations of \one\ 
after fitting a coherent timing solution using a constant
period model, yielding $P=424.1307512(40)$~ms.
}
\label{phaseplot}
\end{figure}

Using the reprocessed data, we iteratively measured the period and
phase of adjacent observations of the 2005 data by the $Z_1^2$ method.
Starting with the the longest observation of 2005 July 11, we
extrapolated the resulting period to the flanking observations,
verifying that the predicted phase and its uncertainty matched to $<
0.1$ cycles the actual phase derived from the adjacent observations.
After completing this procedure for all the 2005 observations a
coherent fit was obtained for the entire set.  Figure~\ref{phaseplot}
shows the phase residuals of the individual observations from the best
fit, which demonstrates the validity of the solution.  The best-fit
period has $Z_1^2 = 412$ and agrees with the period found above from
the incoherent analysis of all observations, while the next highest
peak in the power spectrum has $Z_1^2 = 275$ and is clearly an alias.
To test for a $\dot P$, we then performed a $Z_1^2$ search on a
two-dimensional grid of $P$ and $\dot P$.  The extra parameter did not
increase the peak $Z_1^2$ significantly, meaning no detection of $\dot
P$.

\begin{deluxetable}{ll}
\tablecaption{Spin Parameters of \one }
\tablehead{
\colhead{Parameter} & \colhead{Value}
}
\startdata
Right ascension, R.A. (J2000)\tablenotemark{a} & $12^{\rm h}10^{\rm m}00^{\rm s}\!.91$ \\
Declination, Decl. (J2000)\tablenotemark{a}    & $-52^{\circ}26^{\prime}28^{\prime\prime}\!.4$ \\
Epoch (MJD)                                    & 53562 \\
Spin period, $P$~(s)                            & 0.424130751(4) \\
Period derivative, $\dot P$                    & $(9.6 \pm 9.4) \times 10^{-17}$ \\
Valid range of dates (MJD)                     & 51549--53582 \\
Surface dipole magnetic field, $B_p$~(G)\tablenotemark{b}      & $< 3.5 \times 10^{11}$ \\
Spin-down luminosity, $\dot E$ (ergs s$^{-1}$)\tablenotemark{b} & $< 1.5 \times 10^{32}$ \\
Characteristic age, $\tau_c$ (Myr)\tablenotemark{b}            & $> 24$\\
\enddata
\tablenotetext{a}{\footnotesize Measured from \chandra\ ACIS-I ObsID 3913,
in agreement with \citet{wan07}.}
\tablenotetext{b}{\footnotesize Quantity derived from $2\sigma$ upper limit
on $\dot P$.}
\label{ephemeris}
\end{deluxetable}

The two \xmm\ observations of 2002 August 4 and 6 were easily joined,
resulting in the period listed in Table~\ref{logtable}.  Now knowing
the precise and consistent values of $P$ in 2002 and 2005, we were
able to make a phase-connected combination of the 2001 December \xmm\
observation and the 2002 January \chandra\ one, which are 13 days
apart, by finding an exact period match for the correct peak from
among nearby aliases.  Finally, we combined the set of two \chandra\
observations spanning 2003 June 10--19, which again yielded a
consistent period at the highest peak in the $Z_1^2$
periodogram. After making these coherent combinations, it was not
possible to achieve a further phase-connected solution over a longer
time span, as the intervening cycle counts could not be determined
uniquely.  Therefore, we made a final linear least-squares fit to the
five points listed in Table~\ref{logtable}, yielding the ephemeris
presented in Table~\ref{ephemeris}. This piecewise coherent
measurement, yielding a $2\sigma$ upper limit of $\dot P < 2.8 \times
10^{-16}$, is more than an order of magnitude more precise than the
fully incoherent analysis (see Fig.~\ref{periodplot}).

\section{Interpretation}

Contrary to previous claims, the timing behavior of \one\ does not
require glitches, a binary companion, and perhaps not even accretion
of fallback material, although the latter may still be needed to
contribute to its X-ray spectrum and luminosity.  The absence of
detectable spin variations is due to a weak dipole magnetic field. In
the dipole spin-down formalism, the $2\sigma$ upper limit on $\dot P$
implies, for an isolated pulsar, an energy loss rate $\dot E =
-I\Omega\dot\Omega = 4\pi^2I\dot P/P^3 < 1.5 \times
10^{32}$~ergs~s$^{-1}$, surface magnetic field strength $B_p =
3.2\times10^{19}\sqrt{P\dot P} < 3.5 \times 10^{11}$~G, and
characteristic age $\tau_c \equiv P/2\dot P > 24$~Myr.  In its spin
properties, \one\ is nearly a twin of another CCO, \psr\
\citep{got05,hal07}.  The next section closely follows the discussion
in \citet{hal07}, which anticipated the present result.

\subsection{Cooling and/or Accreting}

The X-ray luminosity of \one\ is of thermal origin, with $L_{\rm bol}
\approx 2 \times 10^{33}\,(d/2\ {\rm kpc})^2$ ergs~s$^{-1}$
\citep{del04}.  This is much larger than the upper limit on its
spin-down power, $\dot E$, and argues that it is mostly residual
cooling.  However, fits to the spectrum require two blackbody
components; the hotter one, of temperature $kT_{\rm BB} = 0.32$~keV,
has an area of only 0.87~km$^2$ \citep{del04}, which may indicate
heating by accretion onto the polar cap.  The canonical area of the
open-field-line polar cap is $R_{\rm pc} = 2\pi^2R^3/Pc \approx
0.27$~km$^2$.  This is $\sim 30\%$ of the fitted blackbody component,
but accretion may cover a wider area.

The characteristic age $\tau_c > 24$~Myr, compared to the remnant age,
estimated as $7$~kyr with an uncertainty of a factor of 3
\citep{rog88}, requires that pulsar was born spinning at its current
period.  A recent population analysis of radio pulsars favors a wide
distribution of birth periods \citep{fau06}, in which 424~ms would be
typical.  Furthermore, as magnetic field is generated by a turbulent
dynamo whose strength depends on the rotation rate of the
proto-neutron star \citep{tho93}, it is natural that pulsars born
spinning slowly would have the weaker $B$-fields; the model of
\citet{bon06} supports this.

There are no young radio pulsars with $B_p < 10^{11}$~G.  Since \one\
is not necessarily beyond the radio pulsar death line, either
empirical \citep{fau06} or theoretical \citep{che93}, there may be
another reason it is radio quiet.  It is possible that low-level
accretion of SN debris prevents CCOs from becoming radio pulsars for
thousands or even millions of years.  Accretion from a fallback disk
\citep{alp01,shi03,eks05,liu06} was one of the theories considered by
\citet{zav04} to explain the now defunct timing irregularities of
\one.  But accretion may still be needed to account for its
radio-quiet and X-ray-hot properties.

If the magnetic field is weak enough that an accretion disk can
penetrate the light cylinder, the hotter portion of the NS surface in
\one\ can be powered by accretion of only $\dot m \approx 10^{13}$
g~s$^{-1}$. It is required that $B_p < 5 \times 10^{11}$~G for \one\ to be
able to accrete in the propeller regime.  But in this limit, the
pulsar would tend to spin down at a rate that is excluded by
observations,
\begin{displaymath}
\dot P \approx 1.2 \times 10^{-14}\,\mu_{29}^{8/7}\,\dot M_{13}^{3/7}\,
\left({M \over M_{\odot}}\right)^{-2/7}\,I_{45}^{-1}\,
\left({P \over 0.424\ {\rm s}}\right),
\end{displaymath}
where the magnetic moment $\mu = B_p\,R^3/2 \approx 10^{29}B_{p,11}$
G~cm$^3$.  Also, $\dot M$ would have to be greater than $10^{13}$
g~s$^{-1}$, as most of the accreting matter is expelled from the
magnetospheric radius rather than accreted.

However, if $B_p < 2 \times 10^9$~G, then \one\ may
accrete as a ``slow rotator,'' and spin up at a
small rate,
\begin{displaymath}
\dot P \approx -2.1 \times 10^{-17}\,\mu_{27}^{2/7}\,\dot m_{13}^{6/7}\,
\left({M \over M_{\odot}}\right)^{3/7}\,
\left(P \over 0.424\ {\rm s}\right)^2.
\end{displaymath}
In this regime, secular spin-up, and torque noise, which may be of the
same magnitude, are below the sensitivity of the existing
measurements.

While flickering is also an indicator of accretion, we do not have
strong evidence of variability of \one\ ($<1\%$ on month timescales).
Also, upper limits on optical/IR emission from \one\ are comparable to
that expected from a geometrically thin, optically thick disk
accreting at the rate required to account for its X-ray luminosity
\citep{zav04,wan07}.  Therefore, it may be necessary to invoke a
radiatively inefficient flow in order to consider accretion.

\subsection{X-ray Absorption Lines}
 
Broad absorption lines in the soft X-ray spectrum of \one\ are
centered at $0.7$~keV and $1.4$~keV \citep{san02,mer02}, and possibly
at 2.1~keV and 2.8~keV \citep{big03,del04}, although the reality of
the two higher-energy features has been disputed \citep{mor05}.
Proposed absorption mechanisms include electron cyclotron in a weak
($8 \times 10^{10}$~G) magnetic field \citep{big03,del04}, atomic
features from singly ionized helium in a strong ($2 \times 10^{14}$~G)
field \citep{san02,pav05}, and iron \citep{mer02}, or oxygen/neon in a
normal ($10^{12}$~G) field \citep{hai02,mor06}.

Our upper limit, $B_p < 3.5 \times 10^{11}$~G, favors the electron
cyclotron model, for at least one of the lines, over all others that
require stronger fields.  The cyclotron prediction, $8 \times
10^{10}$~G, assumes that 0.7~keV is the fundamental energy $E_c =
1.16(B/10^{11}\,{\rm G})/(1+z)$, where $z$ is the gravitational
redshift. Another solution postulates hydrogenic oxygen for the
0.7~keV kine, while the 1.4~keV line is the cyclotron fundamental
\citep{hai02,mor06}. As these authors pointed out, abundant oxygen
may be accreted from supernova debris. One caveat, however, is that
the magnetic field strength at the NS surface can be larger in places
than the global dipole that determines the spin-down rate.

\subsection{Are CCOs a Class?}

The half dozen radio-quiet CCOs are similar in their X-ray
luminosities, high temperatures, and absence of pulsar wind nebulae.
Therefore, they may comprise a fairly uniform class defined by a weak
magnetic field, which in turn results from slow natal rotation.  If
accreting, a slow initial spin is still unavoidable, since even at
Eddington-limited accretion rates, the spin-up and spin-down time
scales in \S 3.1 and in \citet{hal07} are much longer than the ages of
the remnants.  While we do not have definite evidence of accretion in any
CCO, small $B_p$ and large $P$ both make it possible for a pulsar to
accrete at low rates from a SN debris disk; the large $B_p$ and rapid
spin of young radio pulsars prevents such a disk from penetrating the
light cylinder.  In order to test the general applicability of these
results to the class of CCOs, more sensitives searches for their
pulsations are required.  However, prior null results on all of them
suggest that their pulsed amplitudes are very small.

\section{Conclusions}

A comprehensive analysis of all timing data on the X-ray pulsar \one\
has resolved the dilemma of its mysterious spin properties by
correcting previous errors in data processing and analysis.  It is
simply a low magnetic field NS that has no discernible variation in
spin over 5 years.  If an isolated NS, the upper limit on its
spin-down power is much less than its bolometric X-ray luminosity,
which leaves only internal cooling and/or accretion as possible energy
sources.  In either case, \one\ must have been born with a weak
magnetic field and its long rotation period.  We speculate that these
two parameters are causally related and, projecting from the near
twins \one\ and \psr, possibly the physical basis of the CCO class.

An additional benefit of this solution is the new constraint on
proposed absorption-line models for \one\ that depend on the magnetic
field strength.  Our upper limit on the dipole field is close to the
prediction of the electron cyclotron model for both lines, and
perhaps oxygen for one of the lines.  To actually make a
significant measurement of $B_p$ as small as $8 \times 10^{10}$~G from
dipole spin-down would require a fully phase-coherent timing solution
spanning $\ga 6$~yr, assuming that there is no glitch or other timing
noise.  Such a program would also be sensitive to accretion torques at
the lowest rates predicted here for spin-up.  In case $B_p$ is as
small as $10^9$~G and some of the X-ray luminosity of \one\ is due to
accretion from a debris disk, spin-up could be detected.

\acknowledgments

This investigation is based on observations obtained with \xmm
, an ESA science mission with instruments and contributions directly funded by
ESA Member States and NASA. 
Support for this work was provided by NASA through {\it XMM\/} grant
NNX06AH95G and {\it Chandra} Award SAO GO6-7048X.

\end{document}